# Neural network approaches for solving Schrödinger equation in arbitrary quantum wells

A. Radu[1*], C. A. Duque[2]

[1]Department of Physics, Politehnica University of Bucharest, 313 Splaiul Independenţei, Bucharest, RO-060042, Romania

[2]Grupo de Materia Condensada-UdeA, Instituto de Física, Facultad de Ciencias Exactas y Naturales, Universidad de Antioquia UdeA, Calle 70 No. 52-21, Medellín, Colombia

∗Corresponding author: adrian.radu@physics.pub.ro

## Abstract

In this work we approach the Schrödinger equation in quantum wells with arbitrary potentials, using the machine learning technique. Two neural networks with different architectures are proposed and trained using a set of potentials, energies, and wave functions previously generated with the classical finite element method. Three accuracy indicators have been proposed for testing the estimates given by the neural networks. The networks are trained by the gradient descent method and the training validation is done with respect to a large training data set. The two networks are then tested for two different potential data sets and the results are compared. Several cases with analytical potential have also been solved.

Keywords: artificial intelligence, neural network, machine learning, quantum well, Schrödinger equation, numerical solution.

## 1. Introduction

In recent decades, more and more types of problems have been systematically addressed through artificial intelligence methods [1]. In particular, the application of machine learning (ML) to solve problems already admitting conventional solutions can be equally instructive and promising for further extensions to more advanced contexts. A problem of very recent interest, mostly for computational material science and quantum chemistry, is solving the Schrödinger equation (SE) using neural models. So far there are rather few works dealing with this topic, the content of the most noticeable from a physicist point of view being succinctly described in the following.

In a pioneering work [2], M. Sugawara introduced a new methodology for solving the one-dimensional (1D) SE by mixing a genetic algorithm (GA) and a neural network (NN). Approximations of the eigenfunction and eigenvalue were generated by applying an original version of the GA and were later refined by training the NN. A random evaluation point method was used to speed up the convergence in early generations of approximated eigenfunctions. For testing and validation, the method was applied to the harmonic oscillator and Morse potential systems. The acknowledged benefits of this methodology were the relatively small number of unknown parameters and the intrinsically parallel character of the algorithms. K. Mills *et al.* proposed a deep learning method for solving the two-dimensional (2D) SE [3]. They trained a NN to predict the ground and the first excited state energies of an electron in different classes of 2D confining potentials. Their work was intended to generally demonstrate the ability of a properly trained deep NN to rapidly fit the solution to a partial differential equation. A comparison of ML approaches – kernel ridge regression, random forests, and artificial

NNs – was also included in the paper. J. Han *et al.* solved the many-electron SE using deep NNs [4]. They proposed a new family of trial wave functions (WFs), the physical nature of which was explicitly imposed by the Pauli Exclusion Principle. The WF optimization was done through a variational Monte Carlo approach. Successful results were presented for a series of atomic systems: $H_2$, He, LiH, Be, B, and a chain of 10 hydrogen atoms under open boundary conditions. A millihartree accuracy was achieved for two-electron systems. However, a decrease of the accuracy was found as the number of electrons increased. S. Manzhos published a survey of the recent ML techniques used to solve the electronic and vibrational SEs, typically related to computational chemistry [5]. He pointed out that, for abundant and qualitative training data, single-hidden layer NNs seem to be advantageous over multi-layer NNs. Some disadvantages were also mentioned: failing to extrapolate outside of the learning domain, ignoring the symmetries, and costly optimization of parameters. J. Hermann *et al.* proposed a deep NN representation of electronic WFs for molecules with up to 30 electrons and showed that it can outperform other quantum chemistry methods based on variational approaches [6]. H. Li *et al.* used a NN model to solve the static SE by computing multiple excited states and orthogonalized WFs [7]. They discussed the efficiency and scalability of the solver by analyzing 1D and 2D examples of quantum systems. L. Grubišić *et al.* studied two types of NN architectures for approximating eigenmodes localized by a confining potential [8]. A dense deep network was used for a compressed approximation of the ground state and a fully convolutional NN architecture was considered for dealing with the mapping nonlinearly connecting a mesh sample of the potential with the related landscape function. They finally analyzed the accuracy and effectiveness of the proposed methods.

By the present work we intend to approach the Schrödinger problem in quantum wells (QWs) with finite walls and arbitrary potentials, using rather simple NNs. When SE is solved by conventional numerical methods, the computation of the energy eigenvalues and eigenfunctions cannot be conceived as separate, independent problems. The NN approach has the particular advantage of allowing the calculation to be decoupled into two disconnected problems: the estimation of the energies and the estimation of the wave functions. For the energy eigenvalue estimation, an independent NN can be used similarly to how the value of the wave function is estimated by a NN at a particular point in space. The novelty of our paper consists in bringing into discussion and comparing different neural architectures. Given that solving SE does not imply a single numerical estimate, but rather an unknown function discreetly represented as a collection of numbers, two types of NNs with different structures will be proposed for generating these numbers. The NNs are differently trained using a set of potentials, energies, and WFs previously generated by the finite element method (FEM). The main difference between the two NNs used is that the first can be understood as a set of separately trained subnets, for each element of the position discretization. This makes the training process more transparent than for the second network, the latter acting more like a mathematical black box. Three accuracy indicators will be proposed for testing the NNs. The networks are trained by the gradient descent (GD) method using different data batches, and the training is validated with respect to a training data set (DS) containing 151 kilosamples. Then both networks will be tested for two different DSs. Several cases with analytical potential will also be solved and analyzed.

The paper is structured as follows: Section 2 presents the principles of the methods used and Section 3 details the calculations, the results obtained, and the discussions on their account. The conclusions are formulated in Section 4, and the Appendix at the end contains technical details about the sets of potentials used in the calculations.

## 2. Methods

### 2.1. Schrödinger equation and data sets

We consider the SE of a particle confined in a QW:

$$-\frac{\hbar^2}{2m^*}\frac{d^2\varphi(x)}{dx^2} + V(x)\varphi(x) = E\varphi(x), \tag{1}$$

where the finite-walls confinement potential:

$$V(x) = \begin{cases} V_i(x), x \in (-a, a) \\ V_o, x \in [-b, -a] \cup [a, b] \end{cases} \tag{2}$$

is defined so that $0 \leq V_i(x) \leq V_o$ and $V(x)$ has a finite number of discontinuities.

With the notations $\xi \equiv x/a$, $\psi(\xi) \equiv \varphi(a\xi)$, $v(\xi) \equiv V(a\xi)/V_o$, $e \equiv E/V_o$, and $\mu \equiv 2m^*V_o a^2/\hbar^2$, Eq. (1) can be put in the convenient dimensionless form:

$$-\frac{1}{\mu}\frac{d^2\psi(\xi)}{d\xi^2} + v(\xi)\psi(\xi) = e\psi(\xi), \tag{3}$$

where $0 \leq v(\xi) \leq 1$, the WF will be normalized so that $max|\psi(\xi)| = 1$, and $e \in (0,1)$.

We will denote by $\{v_\sigma\}_{1\leq\sigma\leq S}: [-b/a, b/a] \rightarrow [0,1]$ a family of $S$ dimensionless arbitrary confinement functions generated so as to ensure the existence of at least one bound state in each well. A conventional numerical solver is used to calculate the corresponding ground state WFs $\{\psi_\sigma\}$ and energies $\{e_\sigma\}$, with Dirichlet boundary conditions at $\xi = \pm b/a$. The DSs $\{v_\sigma; \psi_\sigma; e_\sigma\}_{1\leq\sigma\leq S}$ thus defined will be later used for training and testing the NNs.

### 2.2. Neural networks architectures and functions

We use two discretizations of the domain on which the dimensionless Schrödinger problem is defined: one for sampling the input data (values of the confinement potential function) and the other for estimating the output data (values of the wave function). Two types of NNs are taken into account; they will be denoted by (a) and (b), as presented in Fig. 1. The input and output discretizations determine the number of neural nodes in the input and output layers, respectively. Both network types have $M$ nodes in the input layer (IL), corresponding to a uniform discretization $\Xi_{in} \equiv \{\xi_i\}_{1\leq i\leq M}$ of the spatial domain $[-b/a, b/a]$, and $P$ nodes in the output layer (OL), corresponding to a different uniform discretization $\Xi_{out} \equiv \{\xi_k\}_{1\leq k\leq P}$ of the same domain. The NNs differ in the architecture of the nodes in the hidden layer (HL): NN(a) has $PN$ neurons separately feeding the $P$ neurons of the OL by individual groups of $N$ neurons each, while NN(b) has $N$ hidden neurons connected to all neurons of the OL. He have practically $P$ similar separate subnets composing the NN(a), each of them predicting the WF at a single point. Conversely, NN(b) cannot be decomposed in separate subnets. Both networks have $PN$ connections between HL and OL, but they differ in the number of connections between IL and HL: $PNM$ for NN(a) and only $NM$ for NN(b).

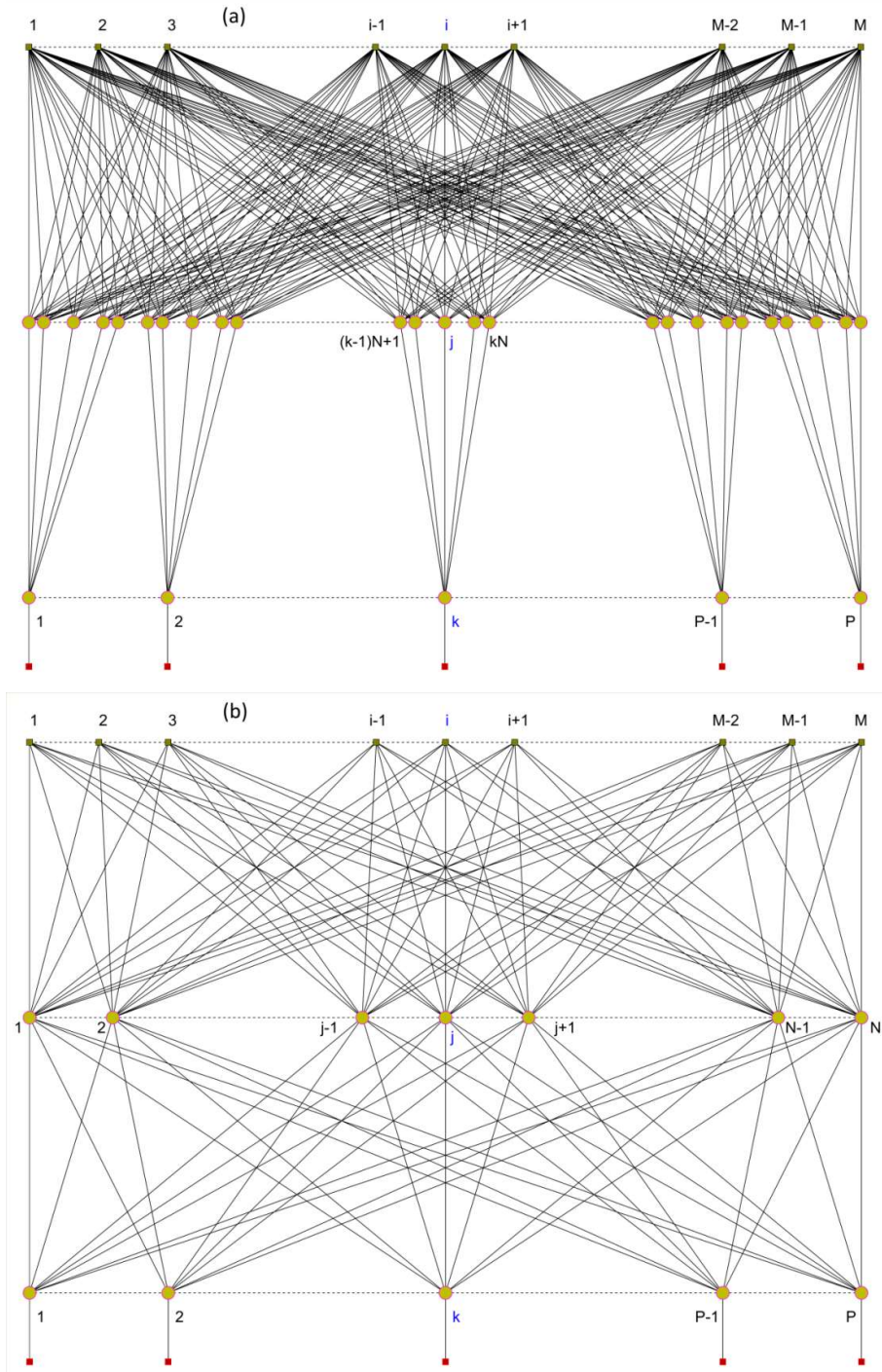

Fig. 1. Connection diagram for: (a) NN composed of $P$ independent subnets, each with different $N$ nodes in the HL and the same $M$ nodes in the IL; (b) NN with common architecture, having $P$ nodes in the OL, $N$ nodes in the HL, and $M$ nodes in the IL.

We consider that a neural activation function nonlinear and continuously differentiable on the entire domain would be the best choice, because this type of function favors an easier calculation and use of the gradients further involved. Also, the dimensionless SE with WF values and energies in the range (0,1) demands an activation function with bounded codomain. The mentioned criteria reduce the possible list to logistic sigmoid and hyperbolic tangent functions. Of these, we prefer the logistic (soft step) function because it has a slightly simpler derivative and it is widely used in the literature.

All neurons have as activation function some variant of the logistic sigmoid. In the following expressions, all matrix operations are element-wise, excepting for the product explicitly noted by "·". The equations below apply to NN(b) and to each subnet in NN(a) (in which case $P$ will be formally 1). Given a particular potential function $v_\sigma$, the data fed by the HL to a neuron in the OL is

$$h_\sigma = H \left\{ \frac{2}{1+\exp[-\Lambda_{\mathrm{h}} \cdot v_\sigma(\Xi_{\mathrm{in}})]} - 1 \right\}, \tag{4}$$

where $h_\sigma$ is an $(N \times 1)$ vector, $H$ is a scale coefficient, $\Lambda_{\mathrm{h}}$ is the $(N \times M)$ weights matrix of the HL, and $v_\sigma(\Xi_{\mathrm{in}})$ is the $(M \times 1)$ vector of the potential data in the IL. The output of the NN is then given by

$$\tilde{\psi}_\sigma = \frac{Q}{1+\exp(-\Lambda_{\mathrm{o}} \cdot h_\sigma)}, \tag{5a}$$

where $\tilde{\psi}_\sigma$ is the $(P \times 1)$ vector of the current estimated WF solution in $\Xi_{\mathrm{out}}$, $Q$ is a scale coefficient, and $\Lambda_{\mathrm{o}}$ is the $(P \times N)$ weights matrix of the OL.

For the energy eigenvalue calculation, an extra subnet of the type in NN(a) can be used, formally replacing $\tilde{\psi}_\sigma$ by $\tilde{e}$ in Eq. (5a), and making $\Lambda_{\mathrm{o}}$ to be an $(1 \times N)$ weights vector:

$$\tilde{e} = \frac{Q'}{1+\exp(-\Lambda_{\mathrm{o}} \cdot h_\sigma)}, \tag{5b}$$

where $Q'$ is the scale coefficient of the energy.

## 2.3. Weights optimization and training

The $(P \times 1)$ loss vector corresponding to the training DS is calculated as:

$$L = \sum_{\sigma=1}^{S} \left[ \tilde{\psi}_\sigma(\Xi_{\mathrm{out}}) - \psi_\sigma(\Xi_{\mathrm{out}}) \right]^2, \tag{6}$$

where $\psi_\sigma(\Xi_{\mathrm{out}})$ is the $(P \times 1)$ vector of the expected solution for the current sample data.

The loss function is defined as:

$$\mathcal{L} = \sum_{k=1}^{P} L_k, \tag{7}$$

and its gradient components with respect to the weights are the $(N \times M)$ and $(P \times N)$ matrices, respectively:

$$\nabla_{\mathrm{h}} \mathcal{L} \equiv \left[ \frac{\partial \mathcal{L}}{\partial \Lambda_{\mathrm{h}}(\mathrm{j,i})} \right]_{\substack{1 \leq \mathrm{j} \leq \mathrm{N} \\ 1 \leq \mathrm{i} \leq \mathrm{M}}}; \tag{8a}$$

$$\nabla_{\mathrm{o}} \mathcal{L} \equiv \left[ \frac{\partial \mathcal{L}}{\partial \Lambda_{\mathrm{o}}(\mathrm{k,j})} \right]_{\substack{1 \leq \mathrm{k} \leq \mathrm{P} \\ 1 \leq \mathrm{j} \leq \mathrm{N}}}. \tag{8b}$$

The starting values of the weights $\Lambda_{\mathrm{h}}^0$ and $\Lambda_{\mathrm{o}}^0$ are randomly chosen and give an initial value $\mathcal{L}^0$ of the loss function. The weights matrices and implicitly the loss function are then iteratively updated by a first order approximation such as to minimize the loss in Eq. (7), i.e. the GD method:

$$\Lambda_{\mathrm{h}}^{\tau+1} = \Lambda_{\mathrm{h}}^{\tau} - \lambda \nabla_{\mathrm{h}} \mathcal{L}^{\tau}; \tag{9a}$$

$$\Lambda_{\mathrm{o}}^{\tau+1} = \Lambda_{\mathrm{o}}^{\tau} - \lambda \nabla_{\mathrm{o}} \mathcal{L}^{\tau}. \tag{9b}$$

Here $0 \leq \tau \leq T$, with $T$ the maximum number of iterations, and $\lambda$ is the learning rate.

The training procedure of the energy subnet is analogous, formally replacing in Eq. (6) $\psi_\sigma$ and $\tilde{\psi}_\sigma$ by $e$ and $\tilde{e}$, respectively, and considering $P$=1 in writing the matrices.

### 2.4. Testing the networks

Three indicators for testing the trained NNs are proposed, based on the dispersion from the expected values in the testing DSs of the predicted energy, WF, and average position of the particle. The relative deviation of the energy is:

$$r_\sigma = \frac{\tilde{e}_\sigma}{e_\sigma} - 1. \tag{10}$$

Comparing the predicted and expected WFs can be done by calculating the relative Euclidean "distance" between $\tilde{\psi}_\sigma$ and $\psi_\sigma$:

$$dw_\sigma \equiv \left\{ \frac{\sum_{k=1}^{P} [\tilde{\psi}_\sigma(\xi_k) - \psi_\sigma(\xi_k)]^2}{\sum_{k=1}^{P} \psi_\sigma^2(\xi_k)} \right\}^{1/2}. \tag{11}$$

Average positions will be compared with:

$$dp_\sigma \equiv \frac{\sum_{k=1}^{P} \xi_k \tilde{\psi}_\sigma^2(\xi_k)}{\sum_{k=1}^{P} \tilde{\psi}_\sigma^2(\xi_k)} - \frac{\sum_{k=1}^{P} \xi_k \psi_\sigma^2(\xi_k)}{\sum_{k=1}^{P} \psi_\sigma^2(\xi_k)}. \tag{12}$$

## 3. Calculations and Results

To train and test the NNs, three DSs have been prepared (hereinafter referred to as DS1, DS2, and DS3). DS1 was used for training and validation and contains 151000 (151k) samples (potentials, WFs, and energies), while DS2 and DS3 were used for testing and contain 14k samples each. Different algorithms have been used to generate the arbitrary confinement potentials corresponding to the three sets, as explained in the Appendix. For computing the expected ground state WFs and energies, the FEM was used to solve the SE with $b/a = 3$ and $\mu = 20$ for all the potential functions in all DSs [9]. Given that for any bound state the WF decreases exponentially to zero far enough away from the QW, the Dirichlet boundary conditions $\psi(\pm b/a) = 0$ were assumed. The chosen value of the parameter $b/a$ was high enough to ensure an accuracy better than one part in a million regarding the calculation of the ground state energy. The estimation of the typical error was made by comparing the FEM result with the exact solution, in the particular case of a square quantum well and for the user-controlled discretization. When solving a 1D differential problem, a common FEM software considers the standard physics-controlled mesh with 101 nodes to be "extremely fine". Indeed, for most cases of usual confinement potentials, the spatial element size of this mesh is small enough that the accuracy of solving the SE is fully sufficient. However, in this work we have allowed some random confinement functions from the DSs to have less trivial behaviors, with faster variations and sometimes with several points of discontinuity. For this generality reason we opted for a user-controlled mesh with more elements. We have thus assumed a larger number of nodes in the IL, ie $M = 1501$, which corresponds to a spatial element of 0.004 for $b/a = 3$. Regarding the spatial discretization used for the WF, corresponding to the OL, this was less fine because the solution of the SE is generally slowly variable with position, even for potentials with fast or frequent changes. The chosen number of nodes in the OL, ie $P = 301$, defined a discretization interval of 0.02, so five times larger than for the IL. In the absence of a systematical optimization study that would provide the minimal settings of the NNs, we preferred to use comfortably large numbers of nodes. The heuristic optimization of the NN architecture is rather difficult due to the huge time consumption required for repeated calculations with different combinations of node numbers. Concerning the neurons in the HL, we found no conclusive or methodical rule in the literature for setting the optimal number. We therefore followed a rather empirical choice of taking it close to the geometric mean of the IL and OL numbers of nodes.

Table 1 contains all numerical parameters and details concerning the architecture and the training/testing of the NNs.

Table 1. Technical aspects of the NNs

| Parameter | NN(a) | NN(b) | Technical details |
|---|---|---|---|
| S | 5k ×P | 14k | Each of the $P$ NN(a) subnets was trained with a different 5k data batch chosen at random from the 151k samples available in DS1. NN(b) was trained with the first 14k data samples in DS1. |
| $M$ | 1501 | | For both architectures we set the same IL and OL. The refinements of the two layers differ because, while the potential can have fast variations, the WF has relatively slow variations. |
| $P$ | 301 | | |
| $N$ | 37 | 672 | There is no methodical rule for the HL optimal size. Numbers were empirically chosen close to the geometric mean of the IL and OL numbers of nodes (for NN(a) subnet and NN(b)). |
| HL neurons | 11137 | 672 | NN(a) and NN(b) have a total of 16727774 and 1210944 connections, respectively. An NN(a) subnet has 55574 connections. |
| $H$ | 5 | | $h(z) = H\left[\dfrac{2}{1+e^{-z}} - 1\right]$ (HL sigmoid – Eq. (4)) |
| $Q$ | 1.1 | | $q(z) = \dfrac{Q}{1+e^{-z}}$ (OL sigmoid – Eqs. (5)) |
| $Q'$ | 1.0 | - | |
| $\Lambda_h^0, \Lambda_o^0$ | | | normally distributed random numbers with mean 0 and standard deviation $10^{-3}$ |
| $\lambda \cdot 10^7$ | 2.5 | 1 | NN(a) has a larger learning rate since its subnets are trained separately, with smaller batches. Larger the batch, smaller the learning rate must be. |
| T | 5000 | | For both NNs this value is sufficient to get cvasistationary weights. |
| Time (a.u.) | 181 | 153 | For NN(a) the given time cumulates the training of all subnets. |

NN(b) looks more like a mathematical black box and the details of the training remain somewhat obscure, in the sense that it is rather unclear why the loss function varies in a certain way over time and what happens to the weight matrices during minimizing this function. Because NN(a) is practically a parallel set of identical $P$ subnets, considerably simpler than NN(b), at least some aspects of training are simpler to perceive. In this sense, it may be useful to visualize in Fig. 2 the statistical manner in which the WFs expected as solutions in the training DS1 are spatially distributed. To build this figure, all 151k WF solutions of the training set were analyzed. The interval $[0,1]$ in which all WFs take values was uniformly discretized, defining a discrete set $\{\omega_m\}_{1\le m\le751}$. For each value $\xi_k$ of the discrete position $\Xi_{out}$ and each value of $m$ between 1 and 750, all the wave solutions from DS1 were counted, which at that position $\xi_k$ take values between $\omega_m$ and $\omega_{m+1}$, the result being denoted by $n_{k,m}$. In this way, a discrete frequency function of two discrete variables is obtained. This frequency can be interpreted in terms of unnormed probability, given the very large number of samples. The decimal logarithm of this function (to which 1 was artificially added to ensure positive values only) is represented in color gradient in the inset of Fig. 2. The results were also represented in Fig. 2 in the form of histogram-like plots with 750 bins whose vertical bars were not drawn, being marked only their heights by discrete points. For seven equidistant particular values from the uniform discretization $\Xi_{out}$ (more precisely for $k = 301, 376, 451, 526, 601, 676,$ and $751$), Fig. 2 shows the logarithmic probability $\lg\left(n_{k,m} + 1\right)$ as a function of the $\omega_m$ variable, in the form of scattered discrete point trails. The trails dispersion along the vertical axis, with respect to some median profile, is produced by statistical fluctuations (being for this reason more visible in the range of low probabilities).

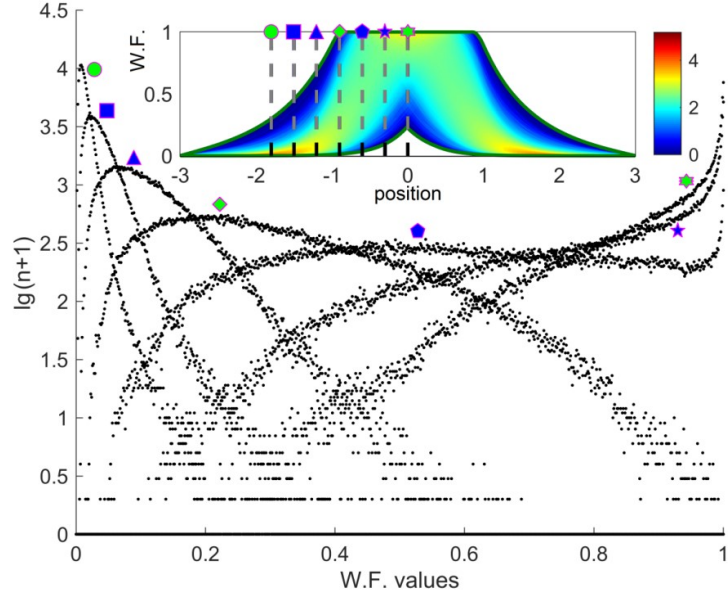

Fig. 2. Logarithmic probability of occurrence of the WF possible values, at several chosen positions, for all 151k samples in DS1. The inset encodes the same type of information in color gradient.

The content of Fig. 2 can be better understood by taking a few examples. In the potential barriers ($|\xi| > 1$), the farther we go from the QW (for example at $\xi_{301} = -1.8$), the more likely that the WF will take values close to zero, and very unlikely to take much higher values. The corresponding trail profile is narrow with a high maximum (see for example the point trail marked with a green circle). In the proximity of the potential walls ($|\xi| \approx 1$) and inside the QW (for example at $\xi_{526} = -0.9$), a wide range of WF values can be taken with comparable probabilities. The trail profile is broader with a moderate maximum (see for example the point trail marked with a green rhombus). At the center of the QW ($\xi_{751} = 0$) the probability is high that the WF takes values very close to the maximum 1, and decreases to small values close to 0.5 and practically to zero near 0.2 (see the point trail marked with a green star). Among the seven discrete point trails in Fig. 2, the one indicated with a blue pentagon (for $\xi_{601} = -0.6$) corresponds to the widest range of almost equiprobable WF values. NN learning process is expected to be slower for the latter position, which will indeed be confirmed by Fig. 3(a).

Figures 3 present the learning curves for the two NNs. It should be noted that each of the 301 subnets of NN(a) was trained with a different batch of 5k samples randomly selected from the 151k samples of the training DS1. NN(b) was trained with a stationary batch composed by the first 14k samples in DS1. The graphs in Fig. 3 show the relative loss function $\mathcal{L}^{\tau}/\mathcal{L}^{0}$ depending on the number $\tau$ of iterations. Figure 3(a) selectively illustrates the results obtained while training the seven individual subnets corresponding to the position values considered when constructing the trails in Fig. 2. It can be seen that learning occurs very quickly (under 100 iterations) for the position marked with a green circle (away from the well): the relative loss function decreases sharply to zero in a few iterations after which it has only a faint variation. The learning also goes fast for the position marked with a green star (the center of the QW), but the value of the first plateau reached is different, considerably higher. In a first stage of 70 iterations a plateau is reached at 12 %, and between 1500 and 3000 iterations, the relative loss function decreases towards a second plateau, at approximately 7.5 %. These observations anticipate poorer post-training prediction accuracy in the center of the QW than in the barrier region, away from the well. The curve marked with a green rhombus (inside the QW, close to the potential wall) indicates a fast learning down to 33 %, followed by a second stage of slower learning, in the first 1000 iterations, down to around 5 %. The value of the relative function after the whole training cycle is about 3.5 %. It is observed that the slowest learning occurs for the position marked with blue pentagon, the one for which in Fig. 2 the maximum range of equiprobable values of the WF was noticed. However, even for this position, the learning is almost complete after the first 2000 iterations, the relative loss function stabilizing at approximately the same value obtained for the center of the QW. After this discussion, it is more apparent that the learning speed of the subnets depends directly on the statistic dispersion of the WF values in the training set, corresponding to the position for which the subnet is defined. Figure 3(b) shows the learning curve of the NN(b), which descends very rapidly in the first 20 iterations, stabilizes at approximately 6-7 % in the first 500 iterations, then decreases again between iterations 500 and 700, and then changing very slowly to a final value below 1 %.

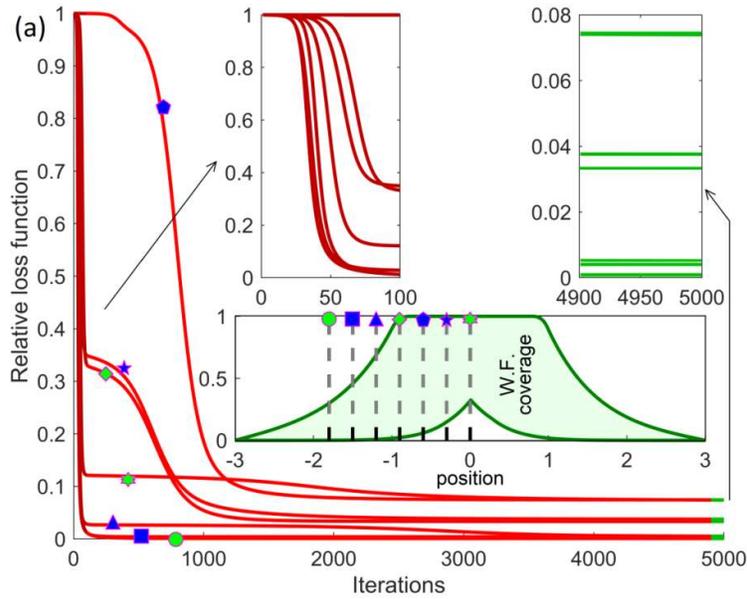

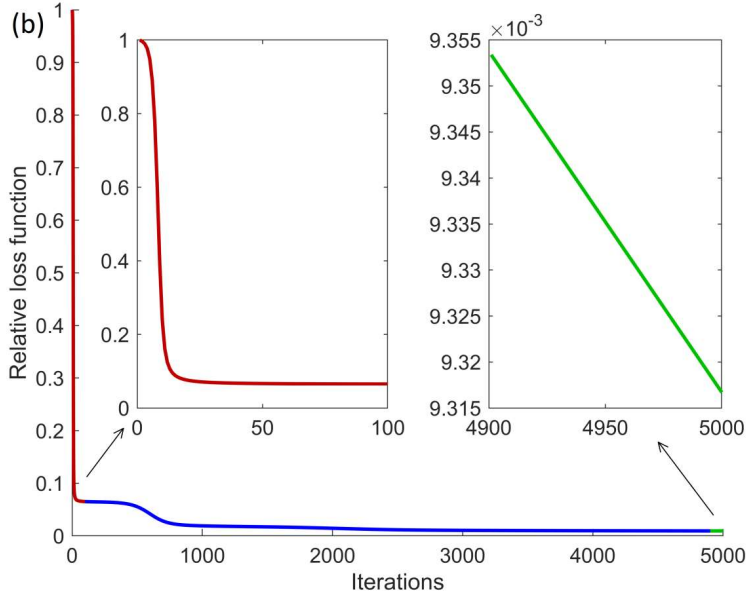

Fig. 3. (a) Learning curves of several subnets in NN(a). The lower inset indicates each subnet position in the spatial domain. The upper left inset details the behavior of the loss curves for the first 100 iterations. The upper right inset indicates the values of the relative loss function at the end of the learning cycle. (b) Learning curve of NN(b). The left/right inset shows the descent of the loss curve for the first/last 100 iterations. For both (a) and (b) plots the discrete points have been connected with solid lines.

To investigate whether the subnets of NN(a) benefited from sufficiently large individual training batches and whether NN(b) does not overfit the 14k training batch, the following validation test is performed. The two completely trained networks are applied on the batch of the first 14k samples from DS1, and then on the entire set of 151k samples, and the results are compared using the effectiveness indicators described in section 2.4. The statistical sets being large, the frequencies of occurrence of different results can be interpreted in the sense of probabilities. If the probability distribution in relation to the range of occurring values is very different between the batch 14k and the entire set 151k, it can be considered that the network training failed. On the contrary, if the results are similar in the limit of small statistical fluctuations, it may be considered that the training was satisfactory. This validation test also serves as a preliminary comparison of the accuracy of the two types of NN. The validation results are illustrated graphically in Figs. 4(a), 4(b) and 4(c), by plotting the normalized probabilities, as functions of the three parameters defined by Eqs. (10), (11), and (12), respectively. The data in Fig. 4(a), obtained with a subnet such as those in NN(a), show a very good agreement between batches 14k and 151k DS1, with statistical fluctuations naturally smaller for the larger DS. It is observed that, on the training set, there is a probability of over 50 % that the subnet will predict the energy with a relative error of less than 5 %, and a probability of almost 90 % to predict the energy within less than 10 % relative error. Lower energy levels are more likely to be predicted with higher relative error, at the same magnitude of the absolute error. It is noted that the maximum probability is reached around the relative error of −5 %, so it is quite likely that the network underestimates the energy by a few percent. Figures 4(b) and 4(c) also demonstrate a very good agreement between the predictions made for training batches 14k and 151k. In addition, these two figures illustrate comparatively the accuracy of the two NNs. There are some differences, especially in terms of the relative deviation of the estimated WF from the expected one. Figure 4(b) shows that the NN(a) predicts the WF with a slightly higher accuracy, as demonstrated by a higher and slightly narrower probability profile. In contrast, Fig. 4(c) seems to indicate a minor advantage of the NN(b) in correctly predicting the average position of the particle in the QW, relative to the samples in DS1. According to these observations, two clear conclusions can be drawn concerning the validation step: (i) the training of both networks is satisfactory, the training batches were large enough; (ii) there are small differences in the accuracy of the two types of network in the predictions made on DS1, with a slight apparent advantage for NN(a). This benefit is added to the fact that the NN(a), being made up of subnets with the same architecture, is more suitable for parallelizing the training calculation.

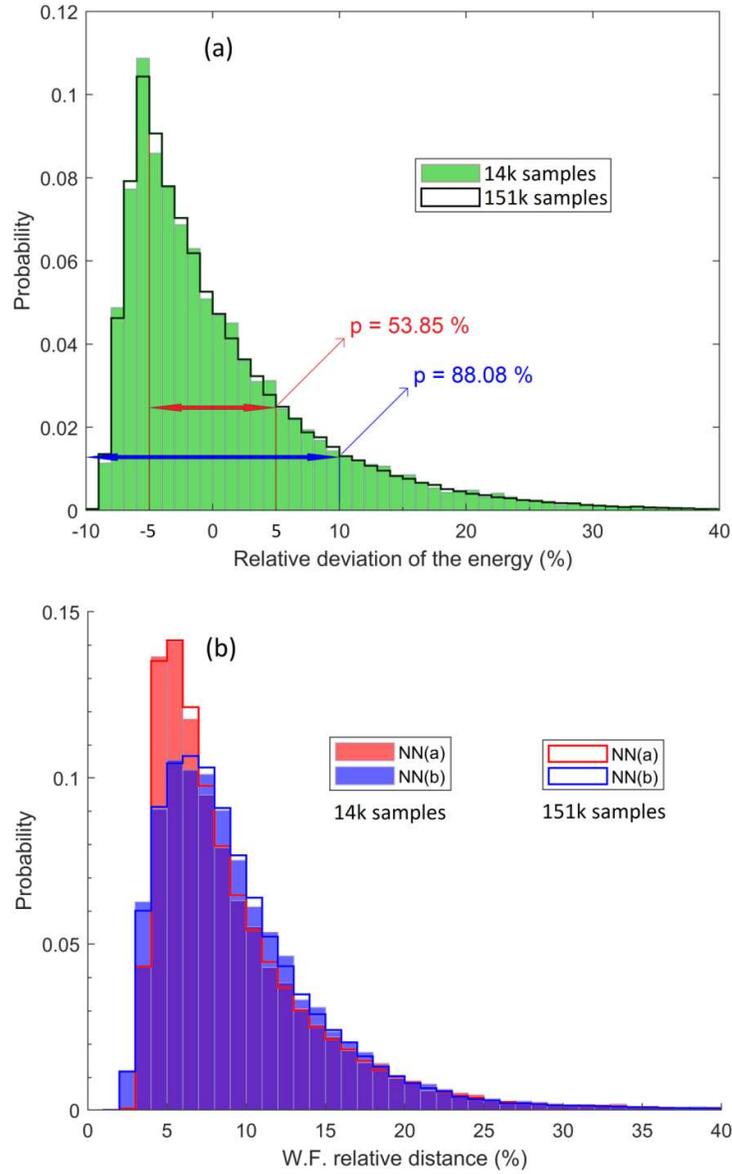

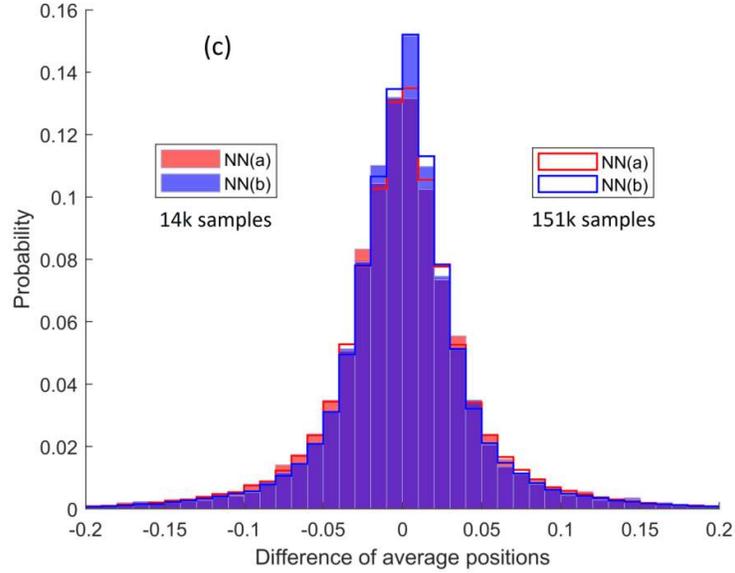

Fig. 4. Multiple histograms comparing the NNs estimations for 14k and 151k samples in the training DS1: (a) relative deviation of energies; (b) relative "distance" between wave solutions; (c) accuracy of average positions. The height of a histogram bar indicates the probability to get a result in the corresponding bin. The bins cover uniformly the range of possible values on the horizontal axis. The face-colored bars are for the 14k samples set and the edge-colored bars are for the entire 151k samples set.

The next stage of the work consists in testing the two NNs, by analyzing their accuracy in estimating the Schrödinger solution obtained by FEM for potential functions from the distinct DS2 and DS3 (14k samples each), which the networks never "met" during training. The same accuracy indicators are used as in Fig. 4 and the results are systematically compared with the standard accuracy obtained on the training with 14k samples from DS1. It should be noted that although different, as shown in the Appendix, the potentials in DS2 are, by the nature of their generation algorithm, somewhat "related" to those in DS1. However, the functions in DS3 are much more different from those in DS1. All probability functions in the test will be normalized such that to make sense of the case comparison. The accuracy measure is a high and narrow probability profile. Comparing the results presented in Figs. 5 shows that, although the accuracy is slightly lower for DS2 and DS3, the NN method still works very well in the new situations. Indeed, the fact that for DS2 the prediction is slightly better than for DS3, may be a sign of the partial influence on the NN of the algorithm model behind the generation of "arbitrary" potentials. Nothing that is based on some algorithm is absolutely "arbitrary", this being different from the pseudo-random nature of the numbers involved. Networks not only "learn" the Schrödinger problem, but also learn something about the particular patterns encountered in the training DSs.

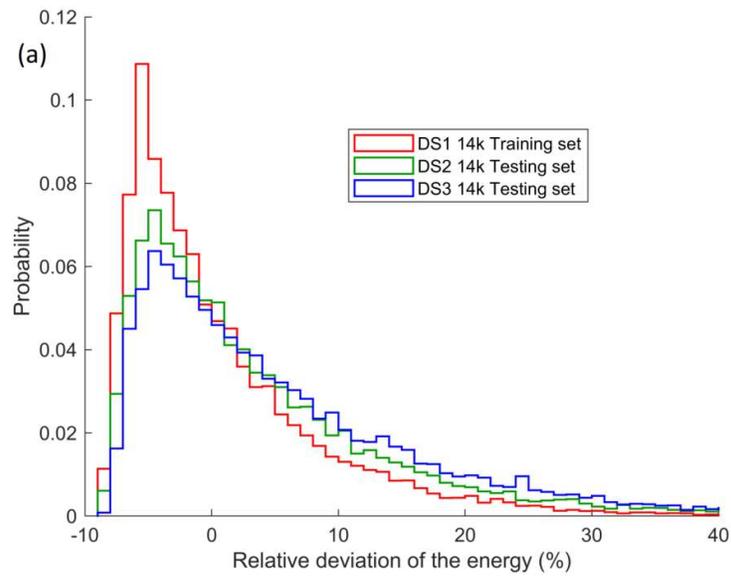

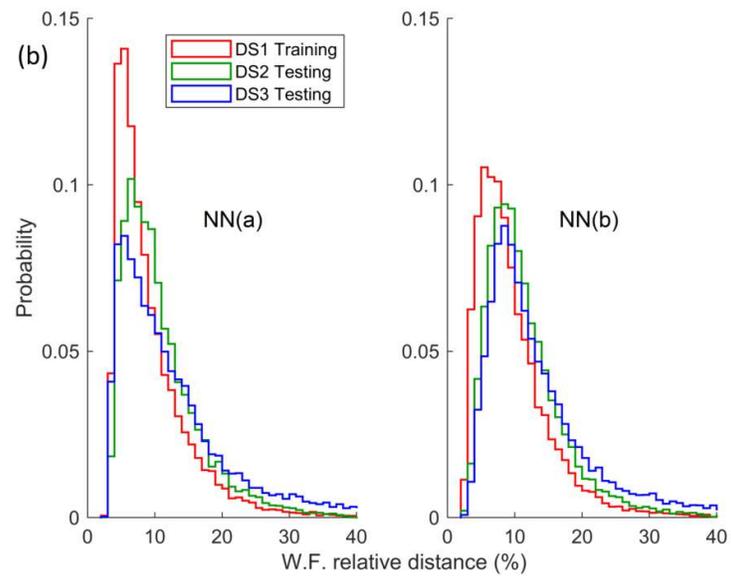

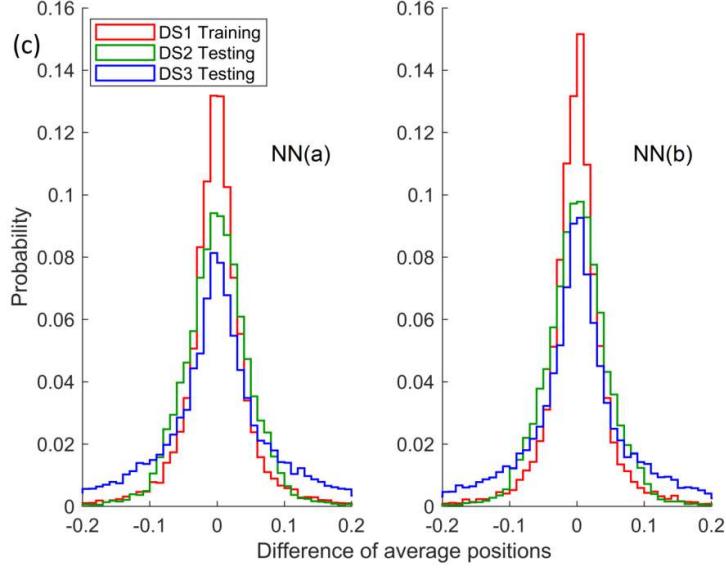

Fig. 5. Multiple histograms comparing the NNs predictions for the 14k samples in each of the testing DS2 and DS3 and the estimations for the 14k samples in the training DS1: (a) relative deviation of energies; (b) relative "distance" between wave functions; (c) accuracy of average positions. The underlying shapes of the probability distributions are indicated by the histogram external perimeters.

Finally, the accuracy indicators of NNs in some particular cases of piecewise analytically-defined confinement potentials (which are not found in any of the three sets of training or testing) are calculated. These cases are also used to plot and compare the aspect of the WFs estimated by each NN with the expected WFs calculated by FEM. Several simple finite-barrier potential types widely used in scientific works on QWs were considered: Square (SQW), Parabolic (PQW), V-shaped (VQW), Trapezoidal (TQW), Semi-parabolic (SPQW), Graded (GQW), Laser-dressed (LDQW), and Asymmetric Step (ASQW). Figures 6(a) - 6(h) display the numerical and graphical results, respectively. In Figs. 6(a), 6(b), 6(c), and 6(g) (QWs with symmetrical potentials) it is found that the accuracy of both NNs is lower in the vicinity of the well center, but very good in the rest of the position interval. For SQW through both NNs a profile of the WF with 2 close maxima is obtained, instead of one. The small fluctuations of the NN(a) solution come from training the subnets with different 5k sample batches at different positions. In Figs. 6(d), 6(e), 6(f), and 6(h) (QWs with asymmetric potentials) a slightly larger deviation of the estimated WF solutions from those expected is observed. The best energy estimate was for the SQW ($r = 1.3$ %), and the weakest for ASQW ($r = 17.9$ %). The WF was best estimated for PQW and VQW, by both networks ($dw \approx 4$ %), and worst for the GQW ($dw \approx 20$ %). The average position of the particle was best estimated for symmetrical QWs, given that the NN solutions were also symmetrical, and the worst for the GQW ($dp \approx 17$ %).

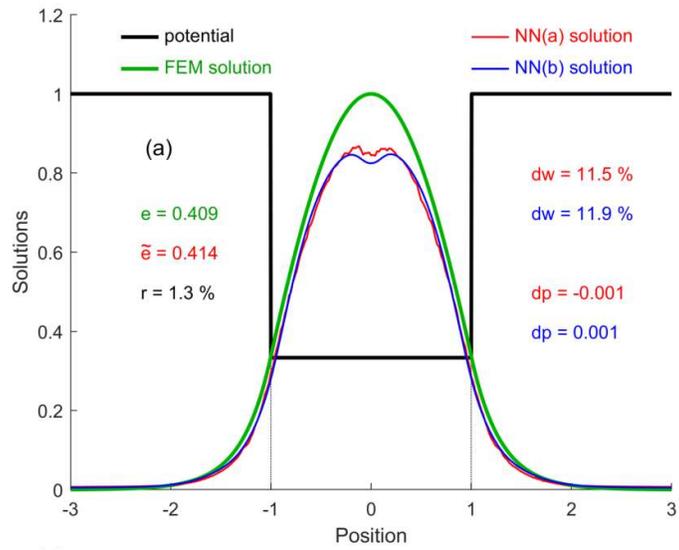

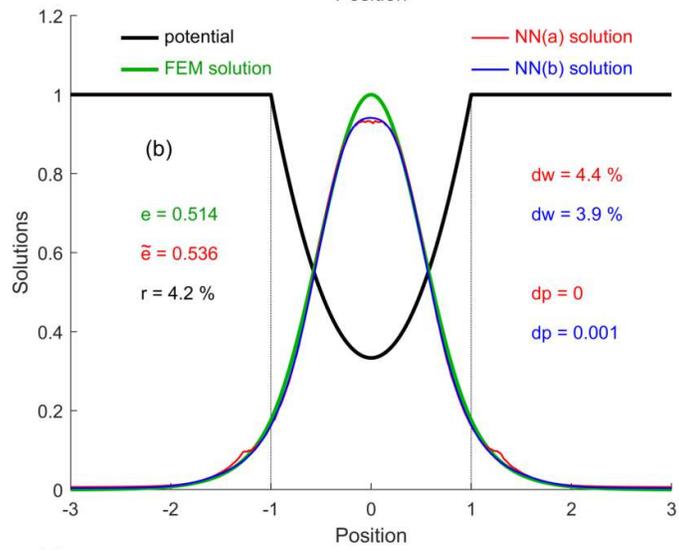

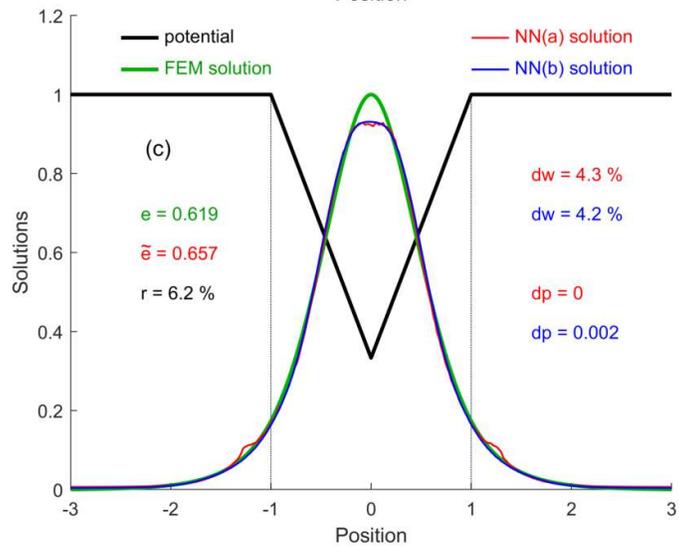

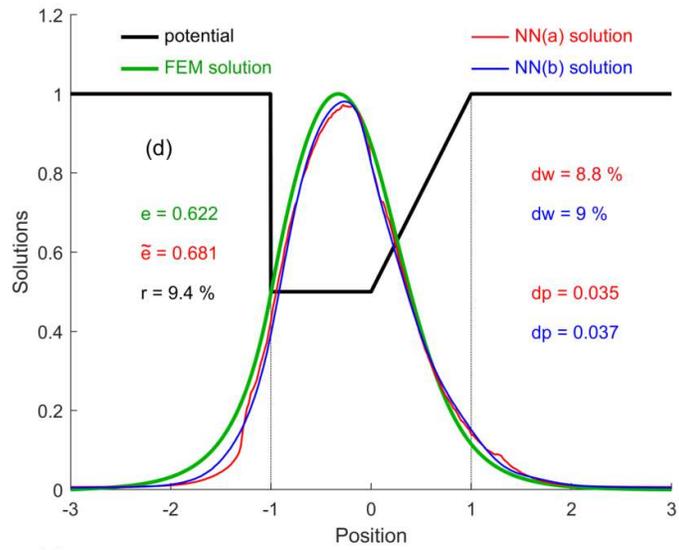

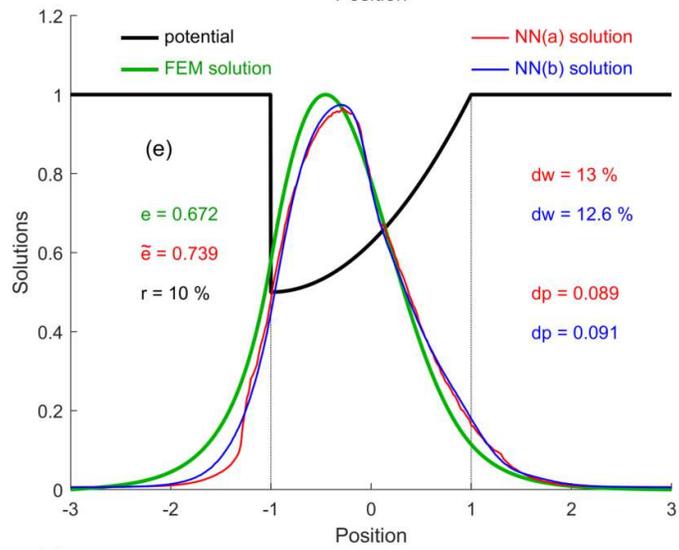

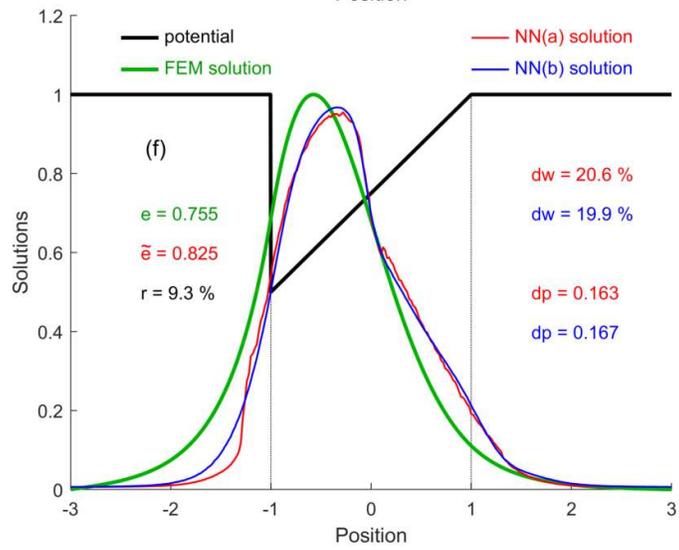

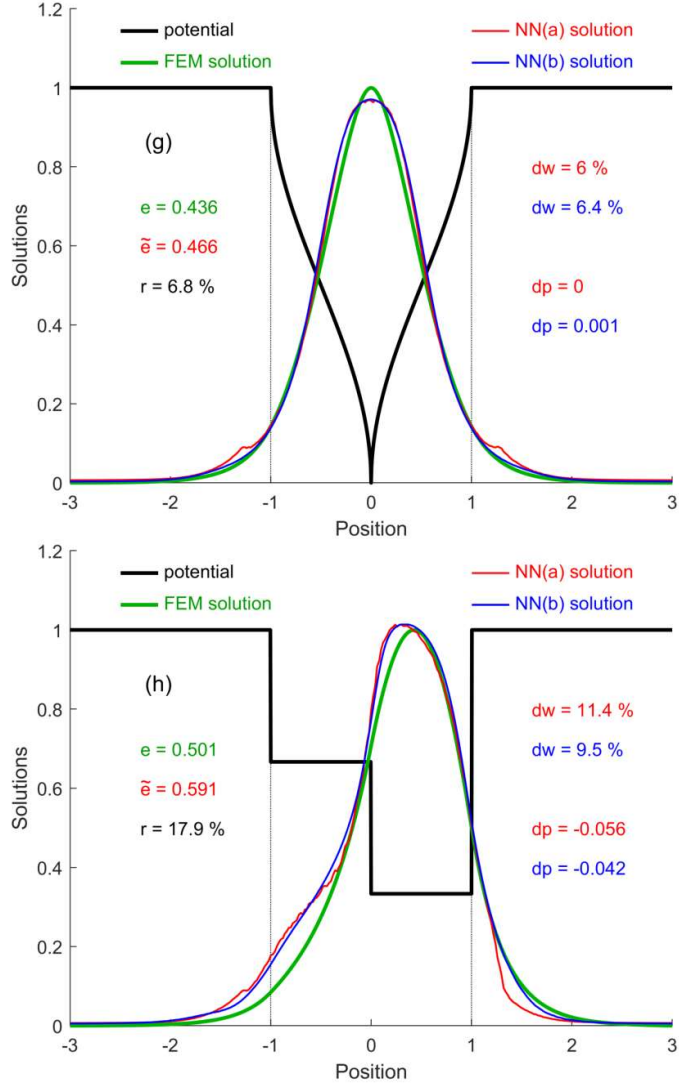

Fig. 6. The FEM and NNs WF solutions and accuracy indicators in 8 particular cases with piecewise analytically-defined potential functions: (a) SQW; (b) PQW; (c) VQW; (d) TQW; (e) SPQW; (f) GQW; (g) LDQW; (h) ASQW.

## 4. Conclusions

In this paper we approached the SE in QWs with finite walls and arbitrary potentials, using the method of ML. Two NNs with different architectures have been proposed and trained using a set of potentials, energies, and WFs previously generated with the classical FEM. The main difference between the two NNs used was that the first can be decomposed into a set of separately trained subnets, for each element of the position discretization. Three accuracy indicators have been proposed for testing the estimates given by the NNs. The networks were trained by the GD method and the training was validated with respect to the training DS containing 151 kilosamples. The learning curves of the two networks were presented and discussed. Then both NNs provided solutions for two different DSs and the results were compared and commented. It was found that the networks have similar effectiveness, slightly lower for the test sets than for the training set. The key advantage of NN(a) is that the calculation is easier to parallelize, while the plus of NN(b) is a smaller number of required neurons and connections. Several cases with analytical potential have also been solved, presenting explicit graphs of estimated WFs. It has been observed that networks have better accuracy in estimating the WFs for potentials with position symmetry.

The improvements and developments that can be made in future works are: (i) using deep networks (with multiple HLs) in order to study whether the accuracy of estimating the WF solutions increases with the number of layers; (ii) using highly mixed training DSs, with arbitrary potential functions generated by multiple different algorithms; (iii) using adaptive learning parameters and/or enhanced optimization algorithms, such as Nesterov or momentum methods; (iv) generalization of NN solutions to 2D and 3D problems (quantum wires and dots); (v) predicting excited energy levels.

## Acknowledgments

CAD is grateful to the Colombian Agencies: CODI-Universidad de Antioquia (Estrategia de Sostenibilidad de la Universidad de Antioquia) and Facultad de Ciencias Exactas y Naturales-Universidad de Antioquia (CAD exclusive dedication project 2021-2022). CAD also acknowledges the financial support from *El Patrimonio Autónomo Fondo Nacional de Financiamiento para la Ciencia, la Tecnología y la Innovación Francisco José de Caldas* (project: CD 111580863338, CT FP80740-173-2019).

## Appendix

1) For each potential function in DS1, a different number $n$ of angular points between 3 and 13 is randomly generated, using a normal distribution with mean 3 and standard deviation 5. Intuitively, this number will be related to the intricacy of the potential spatial variation and is not allowed to be too large, since practical QWs are rather considered. A preliminary discretization $\{\eta_l\}_{1 \leq l \leq n}$ of the well interval $[-1,1]$ is randomly generated using an uniform distribution, and overridden so that $\eta_1 = -1$ and $\eta_n = 1$. An upper envelope discrete function $\{U_l\}_{1 \leq l \leq n}$ is empirically defined in such a way as to ensure the existence of at least one bound energy level in the QW: $U_l = min\left\{\frac{abs(\eta_l)+1}{2}, \frac{7}{10}\right\}$. The discrete potential values $\{u_l\}_{1 \leq l \leq n}$ corresponding to $\{\eta_l\}_{1 \leq l \leq n}$ are then randomly chosen in the interval $[0, U_l]$ by using a uniform distribution, and overridden so that $u_1 = u_n = 1$. Finally, the potential discrete function $\{\xi_i\}_{1 \leq i \leq M} \rightarrow \{v_i\}_{1 \leq i \leq M}$ is obtained by interpolating the set of preliminary points $\{(\eta_l, u_l)\}_{1 \leq l \leq n}$ in the discrete set $\Xi_{in}$ and forcing all the values outside the well to be 1. The interpolation is arbitrarily either linear or piecewise cubic Hermite polynomial. In Fig. A1 some random examples are given.

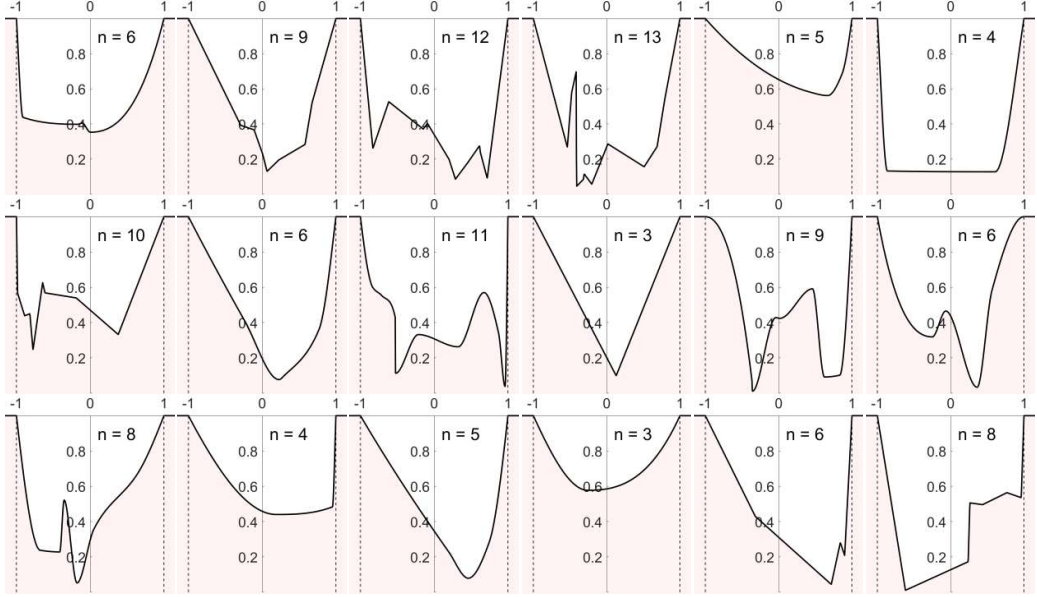

Fig. A1. Several random DS1 potentials.

2) The potentials in DS2 are prepared similarly to those in DS1, except for the final interpolation, which is of the type nearest neighbor (see Fig. A2).

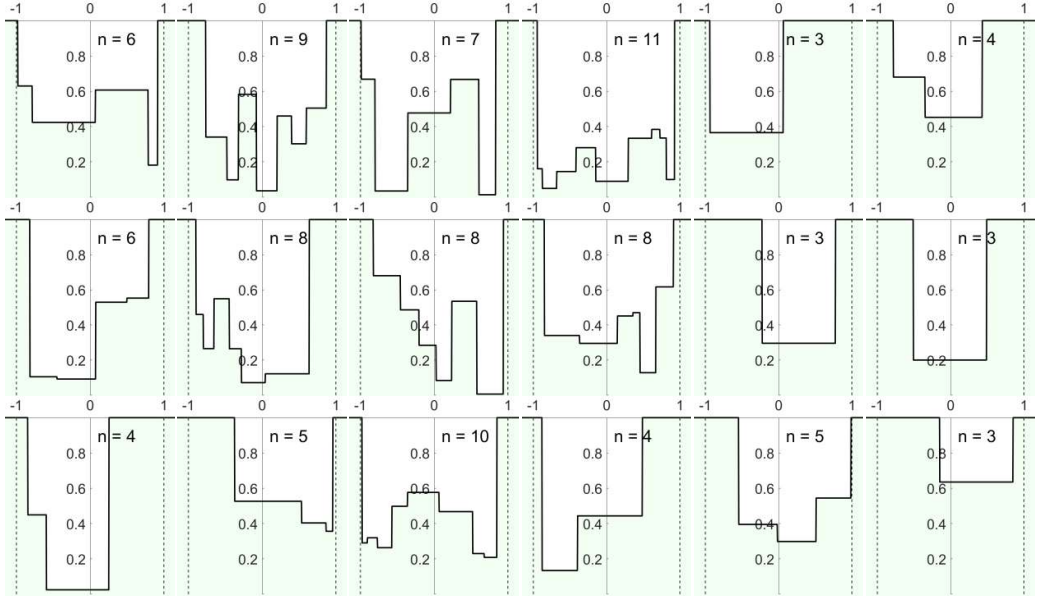

Fig. A2. Several random DS2 potentials.

3) DS3 potentials are based on Fourier synthesis. A random number $n$ of Fourier terms is prepared for each sample following the procedure described for DS1. Fourier coefficients $\{a_l\}_{1 \leq l \leq n}$ and $\{b_l\}_{1 \leq l \leq n}$ are generated randomly in the interval $[-1,1]$ using a uniform distribution. In a first step, the sums of $n$ harmonic terms are calculated: $s_i = \sum_{l=1}^{n}[a_l \cos(l\xi_i) + b_l \sin(l\xi_i)]$, for each position in the discretization $\Xi_{\text{in}}$. Then, an empirical rectification is done in such a way as to ensure at least one bound energy level: $v_i = \frac{7}{10} \cdot \frac{s_i - \min_{1 \leq i \leq M}\{s_i\}}{\max_{1 \leq i \leq M}\{s_i - \min_{1 \leq i \leq M}\{s_i\}\}}$. The potential discrete function $\{\xi_i\}_{1 \leq i \leq M} \to \{v_i\}_{1 \leq i \leq M}$ is then obtained by forcing all the values outside the well to be 1. Examples are given in Fig. A3.

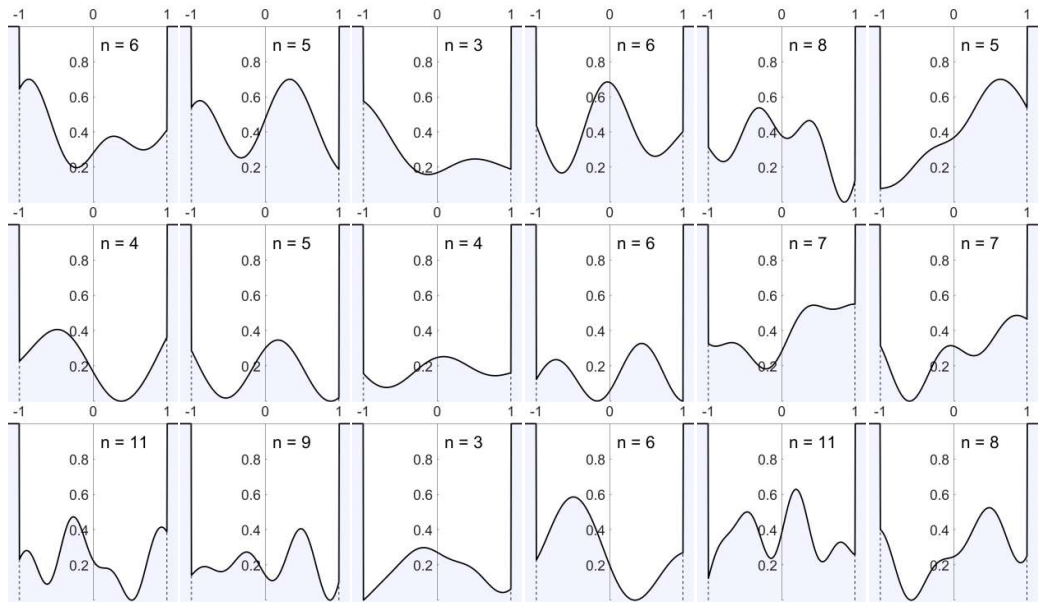

Fig. A3. Several random DS3 potentials.